\documentclass[conference]{IEEEtran}
\IEEEoverridecommandlockouts

\usepackage[letterpaper, left=1in, right=1in, bottom=1in, top=0.75in]{geometry}
\usepackage{cite}
\usepackage{graphicx}
\usepackage{color,soul} 
\usepackage{cite}
\usepackage{amsfonts}
\usepackage{amsmath}
\usepackage{amssymb}   
\usepackage{amsxtra}
\usepackage{amscd}
\usepackage{amsthm}
\usepackage{amsmath}
\usepackage{cite}
\usepackage{mathabx}
\usepackage{cite}
\usepackage{multirow}
\usepackage{hhline}
\usepackage{xcolor}
\usepackage[flushleft]{threeparttable}
\def\BibTeX{{\rm B\kern-.05em{\sc i\kern-.025em b}\kern-.08em
		T\kern-.1667em\lower.7ex\hbox{E}\kern-.125emX}}
	
\begin{document}
%

\title{Deep Convolutional Learning-Aided Detector for Generalized Frequency Division Multiplexing with Index Modulation\\
	{\footnotesize}
}

\author{\IEEEauthorblockN{Merve Turhan}
	\IEEEauthorblockA{Faculty of Electrical and \\Electronics Engineering\\
		Istanbul Technical University\\
		Email: turhanm17@itu.edu.tr\\
		Department of Research and \\Development \\NETAS Telecommunication\\
		Email: mervet@netas.com.tr}
	\and
	\IEEEauthorblockN{Ersin \"{O}zt\"{u}rk}
	\IEEEauthorblockA{Department of Research and \\Development \\NETAS Telecommunication\\
		Email: eozturk@netas.com.tr}
	\and
	\IEEEauthorblockN{Hakan Ali \c{C}{\i}rpan}
	\IEEEauthorblockA{Faculty of Electrical and \\Electronics Engineering\\
		Istanbul Technical University\\
		Email: cirpanh@itu.edu.tr}
	}

\author{Merve Turhan$^{1,2}$, Ersin \"{O}zt\"{u}rk$^2$, Hakan Ali \c{C}{\i}rpan$^1$\\
	$^1$Istanbul Technical University, Faculty of Electrical and Electronics Engineering\\  34469, Maslak, Istanbul, Turkey\\Email: \{turhanm17, cirpanh\}@itu.edu.tr\\
	$^2$Netas, Department of Research and Development\\  34912, Pendik, Istanbul, Turkey\\Email: \{mervet, eozturk\}@netas.com.tr	\vspace*{-0.5cm}}



%


\maketitle

\begin{abstract}
In this paper, a deep convolutional neural network-based symbol detection and demodulation is proposed for generalized frequency division multiplexing with index modulation (GFDM-IM) scheme in order to improve the error performance of the system. The proposed method first pre-processes the received signal by using a zero-forcing (ZF) detector and then uses a neural network consisting of a convolutional neural network (CNN) followed by a fully-connected neural network (FCNN). The FCNN part uses only two fully-connected layers, which can be adapted to yield a trade-off between complexity and bit error rate (BER) performance. This two-stage approach prevents the getting stuck of neural network in a saddle point and enables IM blocks processing independently. It has been demonstrated that the proposed deep convolutional neural network-based detection and demodulation scheme provides better BER performance compared to ZF detector with a reasonable complexity increase. We conclude that non-orthogonal waveforms combined with IM schemes with the help of deep learning is a promising physical layer (PHY) scheme for future wireless networks. 
\end{abstract}

\IEEEpeerreviewmaketitle

\section{Introduction}
The demand for reliable, fast and effective wireless communication methods go on with the growing trend thanks to new applications which have challenging technical requirements. In this sense, orthogonal frequency division multiplexing (OFDM) with multiple numerologies concept has been proposed to meet the requested key performance indicators of fifth generation (5G) wireless networks by Third Generation Partnership Project (3GPP) \cite{3GPP_38201, 3GPP5GNR}. Although OFDM has solid advantages, e.g., simple equalization, robustness to frequency selective fading and easy implementation,  the inabilities of OFDM such as high out-of-band (OOB) emission and high peak-to-average power ratio (PAPR), make it quite disputable to meet the expectations from the physical layer (PHY) of future wireless access technologies \cite{GWunder}. Therefore, improved PHY techniques need to be developed for beyond 5G wireless networks \cite{Zekeriyya1}. 

Generalized frequency division multiplexing (GFDM) \cite{Michailow1} is one of the prominent attempts to cope with the challenges of the future wireless networks. GFDM provides advantages in terms of latency, spectral efficiency, and OOB emission because of block-based structure, reduced overhead of cyclic prefix (CP) and subcarrier-based digitally pulse shaping, respectively. The featured benefit of GFDM is the flexibility that enables time-frequency engineering according to requirements of the target application.

Index modulation (IM) techniques \cite{Basar_IM_NG_IEEE_ACCESS} offer energy and spectral efficiency by utilizing transmission entities to convey digital information innovatively. While spatial modulation (SM) \cite{Mesleh2, Renzo_SM_GenMIMO} utilizes the transmit antennas of a multiple-input multiple-output (MIMO) transmission scheme, OFDM with IM (OFDM-IM) \cite{Ertugrul1,Abualhiga1,Tsonev1} utilizes the subcarrier indices in a multicarrier system to provide alternative ways for transmitting information. Taking account the efficiencies provided by IM, tight integration of GFDM with IM (GFDM-IM) has been considered and innovative transceiver schemes have been introduced \cite{SMGFDMSuboptimal, GFDMIM, GFDMSFIM, gfdm_fim, gfdm_im_ml_sic, smx_gfdm_im}. In \cite{SMGFDMSuboptimal}, the application of the SM-GFDM system has been considered. In \cite{GFDMIM}, the combination of the IM technique with GFDM has been investigated. In \cite{GFDMSFIM}, the combination of GFDM with SM and IM techniques has been considered. In \cite{gfdm_fim}, a GFDM-based flexible IM transceiver, which is capable of generating and decoding various IM schemes has been proposed. In \cite{ gfdm_im_ml_sic}, flexible IM numerology has been proposed to optimize OOB emission, spectral efficiency, and latency jointly. Furthermore, in \cite{ smx_gfdm_im}, a novel MIMO-GFDM scheme, which combines spatial multiplexing (SMX) MIMO transmission, GFDM and IM, has been proposed. Despite having optimized transceiver schemes in terms of OOB emission, spectral and energy efficiency, GFDM-IM schemes suffer high computational complexity with respect to conventional OFDM schemes. 

Deep learning has recently attracted significant attention because of its high performance to solve computationally-burdened problems in various fields such as object detection, natural language processing and computer vision \cite{Goodfellow-et-al-2016}. Considering the unprecedented success of deep learning in classification problems, researchers are eagerly attempting to exploit it for wireless communication. In \cite{Blind_detection}, a pair of blind detectors systems based on the clustering concept has been proposed for SM. In \cite{channel_estimation}, a deep learning-based framework has been presented for channel estimation problem in OFDM. In \cite{Cascadenet}, a zero-forcing (ZF) detector followed by a deep neural network has been proposed for OFDM detection. In \cite{waveform}, a deep complex convolutional network has been developed as an OFDM receiver. 
In \cite{coding} and \cite{AnIntrotoDLPHY}, a communication system has been considered as an autoencoder and communicating binary information through an impaired channel has been treated as reconstruction optimization over impairment layers in a channel autoencoder. This approach has been extended to multi-antenna case in \cite{DLBasedMIMO}. In \cite{Samuel, MIMODetwDL, ModelDriven_DL_MIMO}, deep learning-based MIMO detection schemes have been proposed. Besides, the use of deep learning has also been considered for uplink/downlink channel calibration in massive MIMO systems \cite{Deep_ULDLCalib}. Furthermore, in \cite{DeepIM} and \cite{DLaidedGFDM}, deep learning has been exploited for OFDM-IM and GFDM, respectively. For a comprehensive overview of deep learning aided wireless communication, interested readers are referred to \cite{survey, surveysecond, DLWPHY_OppChal,Zapppone1}. 

In this paper, a novel deep convolutional neural network-based detector is proposed for GFDM-IM scheme in order to reduce the complexity while improving error performance. The proposed detector first applies ZF detector to received signal and then uses a neural network, which is composed of a convolutional neural network (CNN) and a fully-connected neural network (FCNN), to recover the transmitted information from the noisy channel outputs. CNN has three important advantages that can help improve a deep learning model in terms of sparse interactions, parameter sharing, and equivariant representations \cite{Goodfellow-et-al-2016}. The FCNN part has only two fully-connected layers, which can be adapted to yield a trade-off between complexity and bit error ratio (BER) performance. To the best of authors’ knowledge, the proposed method would be the first attempt to exploit a neural network for GFDM-IM detection. Furthermore, a CNN approach is used to detect IM scheme for the first time. We analyze the uncoded BER performance and computational complexity of the proposed detector by comparing with ZF and maximum likelihood (ML) detectors under Rayleigh multipath fading channels. It has been demonstrated that the proposed scheme provides significant BER improvement compared to ZF detector with a reasonable complexity increase. We conclude that non-orthogonal waveforms combined with IM schemes benefiting deep learning is a promising PHY scheme for future wireless networks.

The remaining sections are organized as follows. Section II describes the system model. In Section III, deep convolutional neural network-based joint detection and demodulation (JDD) scheme is presented. Section IV analyzes the computational complexity of the proposed detector. Section V evaluates the BER performance of the proposed scheme with respect to the classical linear detectors. Finally, Section VI concludes the paper.

\section{System Model}
Consider a GFDM symbol with $M$ subsymbols each consisting of $K$ subcarriers, the $m$-th subsymbol is partitioned into $L$ IM blocks, each containing $u=K/L$ subcarrier positions. In an IM block, only $v$ out of $u$ subcarrier positions are selected as active and used to transmit quadrature amplitude modulation (QAM) symbols from $Q$-ary signal constellation $\mathcal{S}$ with $Q$ elements. Thus, an IM block can transmit a $p$-bit binary message $\mathbf{s}_m^l=\left[s_m^l\left(1\right),s_m^l\left(2\right),\ldots,s_m^l\left(p\right)\right]^T$. In each IM block, $p_q=v\log_2(Q)$ bits of incoming $p$-bits sequence are used as QAM-bits. The remaining $p_i=\lfloor\log_2\left(C\left(u,v\right)\right)\rfloor$ bits of this sequence are used to determine the active subcarrier positions. Therefore, we have $\alpha=2^{p_i}$ possible realizations. Here, $C\left(\mu,\nu\right)$ is the binomial coefficient and $\lfloor \cdot \rfloor$ denotes the floor function. Note that active subcarrier positions can be determined using a look-up table or combinatorial methods\cite{Ertugrul1}. As a result, IM blocks  $\mathbf{d}_m^l=\left[d_m^l\left(1\right),d_m^l\left(2\right),\ldots,d_m^l\left(u\right)\right]^T$,
where $d_m^l\left(\gamma\right) \in  \left\lbrace0,\mathcal{S}\right\rbrace$, is constructed according to $p$ input bits \cite{GFDMIM}. Then, IM blocks are first concatenated to obtain the GFDM-IM subsymbol $\mathbf{d}_m=\left[d_{0,m},d_{1,m},\ldots,d_{K-1,m}\right]$ and the resulting GFDM-IM subsymbols are combined to form the GFDM-IM symbol	
\begin{multline*}
\mathbf{d}=\\\left[d_{0,0},\ldots,d_{K-1,0},d_{0,1},\ldots,d_{K-1,1},\ldots,d_{K-1,M-1}\right],
\end{multline*}
where $d_{k,m} \in  \left\lbrace0,\mathcal{S}\right\rbrace$, for $k=0,\ldots,K-1, m=0,\ldots,M-1$, is the data symbol of $k$-th subcarrier on $m$-th subsymbol. After that, the GFDM-IM symbol $\mathbf{d}$ is modulated using a GFDM modulator and the resulting GFDM transmit signal can be expressed as
\begin{equation}
\mathbf{x}=\mathbf{A}{\mathbf{d}},
\label{eq:gfdm_xad}
\end{equation}	
where $\mathbf{A}$ is an $N \times N$ GFDM modulation matrix \cite{Michailow1}, $N=KM$. Finally, a CP with length $N_{\text{CP}}$ is appended to $\mathbf{x}$ and the resulting vector $\tilde{\mathbf{x}}=\left[\mathbf{x}\left(KM-N_{\text{CP}}+1:KM\right)^T, \mathbf{x}^T\right]^T$
is sent over a frequency-selective Rayleigh fading channel.

At the receiver side, assuming that perfect synchronization is ensured, CP is longer than the tap length of the channel $(N_{\textrm{Ch}})$ and the wireless channel remains constant during the transmission of a GFDM symbol,  the received signal vector $\mathbf{y}$ can be expressed as
\begin{equation}
\mathbf{y}=\mathbf{H}\mathbf{x}+\mathbf{n}
\label{eq:gfdm_basic_system_model}
\end{equation}
after the removal of CP. Here, $\mathbf{y}=[y(0), y(1),\ldots,y(N-1)]^{\text{T}}$ is the vector of the received signals, ${\mathbf{H}}$ is the $N \times N$ circular convolution matrix constructed from the channel impulse response coefficients given by $\mathbf{h}=\left[h(1),h(2),\ldots,h(N_{\textrm{Ch}})\right]^\text{T}$, and $\mathbf{n}$ is an $N \times 1$ vector of additive white Gaussian noise (AWGN) samples. The elements of $\mathbf{h}$ and $\mathbf{n}$ follow $\mathcal{CN}(0,1)$ and $\mathcal{CN}(0,\sigma_w^2)$ distributions, respectively, where $\mathcal{CN}(\mu,\sigma^2)$ shows the distribution of a circularly symmetric complex Gaussian random variable with mean $\mu$ and variance $\sigma^2$. After substituting Eq. \ref{eq:gfdm_xad} in Eq. \ref{eq:gfdm_basic_system_model}, we obtain
the equivalent channel of the GFDM-IM scheme as
\begin{equation}
\mathbf{y}=\mathbf{H}\mathbf{A}\mathbf{d}+\mathbf{n}=\widetilde{\mathbf{H}}\mathbf{d}+\mathbf{n}.
\label{eq:gfdm_equivalent_system_model}
\end{equation}

\section{Deep Detection and Demodulation}
\begin{figure*}[!t]
	\centering
	\begin{center}\resizebox*{12cm}{!}{\includegraphics[scale = 0.8]{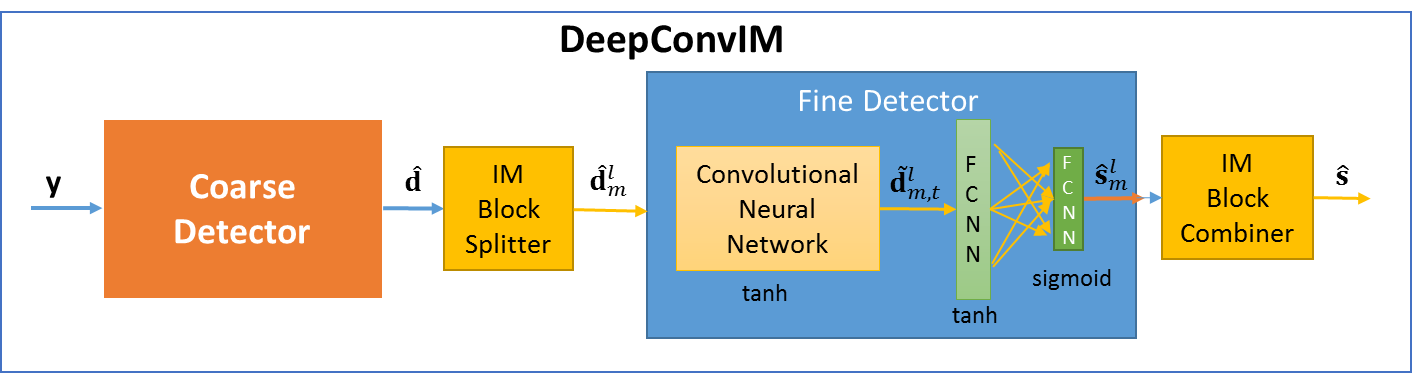}}
		\vspace*{-0.20cm}
		\caption{Block diagram of the deep convolutional neural network-based GFDM-IM detector.}
		\vspace*{-0.20cm}
		\label{fig:dl_jdd}
	\end{center}
	\vspace*{-0.40cm}
\end{figure*}

The block diagram of the proposed deep convolutional neural network-based joint GFDM-IM detection and demodulation scheme, termed as DeepConvIM, is shown in Fig \ref{fig:dl_jdd}. It is assumed that the receiver has the channel information. In contrast to OFDM-IM, GFDM-IM subcarriers can be non-orthogonal to each other due to non-rectangular pulse shaping. Therefore, the inherent ICI prevents the frequency domain decoupling of GFDM-IM subcarriers for both single-input single-output (SISO) and MIMO transmission schemes. As a result, simultaneous detection of all subcarriers is required for optimum decision. Since this process is computationally infeasible, low complexity solutions are required for the optimum detection problem of GFDM-IM. Inspired from \cite{Cascadenet}, the proposed detector has two parts as coarse detector and fine detector. This two stage approach prevents  getting stuck of neural network in a saddle point and enables the processing IM blocks independently. First, coarse detector uses ZF detector in order to process channel and GFDM modulation effects jointly. The output vector of coarse detector can be expressed as
\begin{equation}
\hat{\mathbf{d}}=\left(\widetilde{\mathbf{H}}^{H}\widetilde{\mathbf{H}}\right)^{-1}\widetilde{\mathbf{H}}^{H}\mathbf{y}.
\end{equation}
Since coarse detector operates on the equivalent channel of the GFDM-IM scheme, the remaining parts can handle the IM blocks individually. Therefore, fine detector processes the IM blocks independently. IM Block Splitter partitions the pre-processed received vector $\hat{\mathbf{d}}$ into IM blocks $\mathbf{\hat{d}}_m^l=\left[{\hat{d}}_m^l\left(1\right),{\hat{d}}_m^l\left(2\right),\ldots,{\hat{d}}_m^l\left(u\right)\right]^T$. The fine detector part of DeepConvIM uses a CNN followed by a FCNN, which is expressed as
\begin{equation}
\hat{\bold{s}}_{m}^{l}=\mathbf{\theta{(\hat{d}_{m}^{l})}},
\label{eq:2}
\end{equation}
where $\theta$ represent the total of trainable parameters. The CNN part of the fine detector convolves the IM block $\hat{\mathbf{d}}_{m}^{l}$ with the kernel filter $\bold{a}_{t} = \big[a_{t,R} \quad a_{t,I}\big]$, adds bias $\bold{c}_{t}$, for $t = 1,...T$, with stride $1$, and the modified received IM block can be expressed as
\begin{multline*}
\check{{d}}_{m,t}^{l}(\gamma)=\\ \mbox{tanh}({a_{t,R}*\Re(\hat{d}_{m}^{l}(\gamma))+ a_{t,I}*\Im(\hat{d}_{m}^{l}(\gamma))+c_{t}}),
\label{eq:5}
\end{multline*}
where $\mbox{tanh}$ is an activation function, for $\gamma = 1,...,u$. Here $\bold{a}_{t}$ and $\bold{c}_{t}$ are called convolution trainable parameters. Notice that unlike \cite{DeepIM}, DeepConvIM does not need the energy of the received signal. The FCNN part of the fine detector gets the output of the CNN and performs deep detection by using $\left\{\mathbf{W,b}\right\}$ trainable parameters, where $\mathbf{W} = {\big[\bold{w}_{1} ,\bold{w}_{2}]}
$ contains weights parameters and $\bold{b} = [b_{1}, b_{2}]$ contains bias parameters. That is, The FCNN part uses only two fully-connected layers, hidden layer has $\tau$ nodes the output layer has $p$ nodes as expected. The output of fine detector can be expressed as 
\begin{equation}
\bold{\hat{s}}_{m}^l = \mbox{sigmoid}(\bold{w}_{2}({\mbox{tanh}(\bold{w}_{1}\bold{\check{d}}_{m,t}^{l}+b_{1})})+b_{2},
\label{eq:8}
\end{equation}
where $\mbox{sigmoid}$ is an activation function. Finally, IM Block Combiner combines the output of the fine detector and forms the transmitted information bits.

The aim of the training stage of DeepConvIM is to find $\theta$ parameters in order to minimize the loss function, which is expressed as  $loss(\bold{s}_{m}^l,\bold{\hat{s}}_{m}^l) = \|\bold{s}_{m}^l-\bold{\hat{s}}_{m}^l\|$. Before training, GFDM-IM simulation training data is generated and divided into batchs ($B$). At first, the $\theta$ is randomly initialized. Throughout the training, $\theta$ is updated according to stochastic
gradient descent (SGD) algorithm for every batch, which is expressed as 
\begin{equation}
{\theta_{+} = \theta - \eta\bigtriangledown loss(\bold{s}_{m},\hat{\bold{s}}_{m})},
\end{equation}
where $\eta$ is learning rate.

\section{Complexity Analysis} \label{complexity_analysis}
Computational complexity of ZF, ML and DeepConvIM detectors is investigated from the standpoint of number of complex multiplications (CMs) and given in Table \ref{tab:compucomp_details}. Here, $\Psi_{J\times I}$ and $\Phi_{J\times I}$ are used for $J\times I$ matrices, $\psi_{J\times 1}$ and $\phi_{J\times 1}$ stand for $J\times 1$ vectors. Notice that using complex numbers is not yet supported by any popular deep learning frameworks and FCNN part of DeepConvIM operates on real numbers thanks to CNN part. Since one complex multiplication can be carried out with at least three real multiplications, the number of multiplications belonging to neural networks parts of DeepConvIM are divided to three in order to refer them as complex multiplications. The summary of the results is given in Table \ref{tab:compucomp_sum}. From Table \ref{tab:compucomp_sum}, it is observed that while ZF and ML detectors have the lowest and the highest complexity, respectively, DeepConvIM provides an intermediate solution with regard to computational complexity.

\begin{table*}[!t]
	\begin{center}
		\begin{threeparttable}
			\caption{Computational Complexity of ZF, ML and DeepConvIM Detectors}\vspace*{0.1cm}
			\label{tab:compucomp_details}
			\begin{tabular}[c]{|l||c|c|c|c|} \hline
				\textit{Detector} & \textit{Process} & \textit{Operation} & \textit{Execution Count} & \textit{Complexity (CMs)}  \\ \hline \hline
				\multirow{3}{*}{ZF} & Forming $\widetilde{\mathbf{H}}$ & ${\Phi_{N\times N}\Psi_{N\times N}}^{\dagger}$  & 1 & $N_\text{Ch}N^2$ \\ \hhline{~----}
				& JDD & ${\left({\Phi_{N\times N}}^\text{H}{\Phi_{N\times N}}\right)}^{-1}{\Phi_{N\times N}}^\text{H}\phi_{N\times 1}$ & 1 &  $3N^3+N^2$ \\ \hhline{~----}
				& Decision & $\text{min}\left({\lVert\phi_{u\times 1}-\psi_{u\times1})\rVert}^2\right)$ & $ML$ &  $u\alpha Q^vML$ \\ \hline
				\multirow{2}{*}{ML} & Forming $\widetilde{\mathbf{H}}$ & ${\Phi_{N\times N}\Psi_{N\times N}}^{\dagger}$  & 1 & $N_\text{Ch}N^2$ \\ \hhline{~----}
				& Decision & $\text{min}\left({\lVert\phi_{N\times 1}-\left(\Phi_{N\times N}\psi_{N\times1}\right)\rVert}^2\right)^{\dagger\dagger}$ & $\left(\alpha Q^v\right)^{ML}$ &  $\left(\alpha Q^v\right)^{ML}\left(NvML+N\right)$ \\ \hline
				\multirow{4}{*}{DeepConvIM} & Forming $\widetilde{\mathbf{H}}$ & ${\Phi_{N\times N}\Psi_{N\times N}}^{\dagger}$  & 1 & $N_\text{Ch}N^2$ \\ \hhline{~----}
				& JDD & ${\left({\Phi_{N\times N}}^\text{H}{\Phi_{N\times N}}\right)}^{-1}{\Phi_{N\times N}}^\text{H}\phi_{N\times 1}$ & 1 &  $3N^3+N^2$ \\ \hhline{~----}
				& CNN & ${\left(2uT+uT\lambda\right)/3}^{\dagger\dagger\dagger}$ & $ML$ & $\left(uT\left(2+\lambda\right)\right)ML/3$ \\ \hhline{~----}
				& FCNN & ${\left(uT\tau+\tau\lambda+\tau p+p\delta\right)/3}^{\dagger\dagger\dagger\dagger}$ & $ML$ & $\left(uT\tau+\tau\lambda+\tau p+p\delta\right)ML/3$\\ \hline		
			\end{tabular}
			\begin{tablenotes}
				\small
				\item $^{\dagger}$ In every row of ${\mathbf{H}}$, which is $\Phi$ in this case, only $N_{Ch}$ out of $N$ elements are non-zero.
				\item $^{\dagger\dagger}$ In $\psi$, only $vML$ complex elements are nonzero.
				\item $^{\dagger\dagger\dagger}$ $\lambda$ refers to number of real multiplications required for $tanh$ function.
				\item $^{\dagger\dagger\dagger\dagger}$ $\delta$ and $\tau$ refers to number of real multiplications required for $sigmoid$ function and the number of nodes of the hidden layer of FCNN, respectively.
			\end{tablenotes}
		\end{threeparttable}		
	\end{center}
	\vspace*{-0.4cm}
\end{table*}

\begin{table*}[!t]
	\begin{center}
		\caption{Summary of the Computational Complexity of ZF, ML and DeepConvIM Detectors}\vspace*{-0.20cm}
		\label{tab:compucomp_sum}
		\begin{tabular}[c]{|l||c|} \hline
			\textit{Detector} & \textit{Total Complexity (CMs)} \\ \hline \hline
			ZF & $3N^3+N^2\left(1+N_\text{Ch}\right)+\alpha Q^vML$ \\ \hline
			ML & $\left(\alpha Q^v\right)^{ML}\left(NvML+N\right) +N_\text{Ch}N^2$  \\ \hline
			DeepConvIM & $3N^3+N^2\left(1+N_\text{Ch}\right) + \left(uT\left(2+\lambda\right)\right)ML/3 + \left(uT\tau+\tau\lambda+\tau p+p\delta\right)ML/3$ \\ \hline
		\end{tabular}
	\end{center}
	\vspace*{-0.4cm}
\end{table*}

\section{Numerical Results}

\begin{table}[!t]
	\begin{center}
		\caption{A look-up table example for $u=4,v=2$.}\vspace*{-0.20cm}
		\label{tab:Look_up1}
		\begin{tabular}[c]{|c||c||c|} \hline
			\textit{Bits} & \textit{Indices} & \textit{IM block}  \\ \hline \hline
			$[0\,\,\, 0]$ & $\left\lbrace 1, 2\right\rbrace $ & $\begin{bmatrix}s_{\chi} & s_{\zeta} & 0 & 0   \end{bmatrix}^T$  \\ \hline
			$[0\,\,\, 1]$ & $\left\lbrace 2,3\right\rbrace$ & $\begin{bmatrix}0 & s_{\chi} & s_{\zeta} & 0   \end{bmatrix}^T$ \\ \hline
			$[1\,\,\, 0]$ & $\left\lbrace 3, 4\right\rbrace$ & $\begin{bmatrix} 0 & 0 & s_{\chi} & s_{\zeta}     \end{bmatrix}^T$ \\ \hline
			$[1\,\,\, 1]$ & $\left\lbrace 1,4\right\rbrace$ & $\begin{bmatrix}s_{\chi} & 0 & 0 &  s_{\zeta}   \end{bmatrix}^T$  \\ \hline
		\end{tabular}
	\end{center}
	\vspace*{-0.40cm}
\end{table}

\begin{table*}[!t]
	\begin{center}
		\caption{Fine Detector Model Parameters }\vspace*{-0.20cm}
		\label{tab:model_params}
		\begin{tabular}[c]{|l||c|c|} \hline
			\textit{Description} & \textit{Parameter} & \textit{Value}  \\ \hline \hline
			Number of Kernel Filter (for BPSK transmission) & $T$ & 16  \\ \hline
			Number of Kernel Filter (for 4-QAM transmission) & $T$ & 32  \\ \hline
			Number of Kernel Filter (for 16-QAM transmission) & $T$ & 64  \\ \hline
			Number of Nodes of Hidden Layer (for BPSK transmission) & $\tau$ & 64  \\ \hline
			Number of Nodes of Hidden Layer (for 4-QAM transmission) & $\tau$ & 128  \\ \hline
			Number of Nodes of Hidden Layer (for 16-QAM transmission) & $\tau$ & 256  \\ \hline
			Learning Rate & $lr$ & 0.0008\\ \hline
			Batch Size & $B$ & 1000  \\ \hline
			
		\end{tabular}
	\end{center}
	\vspace*{-0.50cm}
\end{table*}

\begin{table}[!t]
	\begin{center} 
		\caption{Fine Detector  Model Summary}\vspace*{-0.20cm}
		\label{tab:model_summary}
		\begin{tabular}{|c|c|c|}
			\hline
			\textit{Layer} & \textit{Output Shape} & \textit{Activation Func.} \\	
			\hline	
			\hline Input & (B,2,u,1) & None \\ 
			\hline Conv2d & (B,1,u,T) & tanh \\ 
			\hline Flatten & (B,uT) & None \\
			\hline Dense & (B,$\tau$) & tanh \\
			\hline Dense & (B,p) & sigmoid \\
			\hline
		\end{tabular}
	\end{center}
	\vspace*{-0.50cm}
\end{table}

\begin{table*}[!t]
	\begin{center} 
		\caption{The Total Number of CMs for ZF, ML and DeepConvIM Detectors}\vspace*{-0.20cm}
		\label{tab:compucomp_all}
		\begin{tabular}[c]{|l||c|c|c|c|} \hline
			\textit{Configuration} & \textit{ZF} & \textit{DeepConvIM} & \textit{ML} \\ \hline \hline
			BPSK, $K=32, M=1$ & $1.07\times10^{5}$ & $1.18\times10^{5}$ & $2.33\times10^{12}$  \\ \hline
			4-QAM, $K=32, M=1$ & $1.09\times10^{5}$ & $1.56\times10^{5}$ & $1.48\times10^{22}$ \\ \hline
			16-QAM, $K=32, M=1$ & $ 2.37\times10^{5}$ & $4.20\times10^{5}$ & $ 4.15\times10^{36}$ \\ \hline
			BPSK, $K=8, M=1$ & $2.08\times10^{3}$ & $5.15\times10^{3}$ & $1.07\times10^{4}$  \\
			\hline
			BPSK, $K=8, M=3$ & $4.61\times10^{4}$ & $5.54\times10^{4} $ & $5.23\times10^{9}$ \\
			\hline
		\end{tabular}
	\end{center}
	\vspace*{-0.5cm}
\end{table*}

In this section, the BER performance of DeepConvIM has been compared to ZF and ML detection methods by computer simulations for Rayleigh fading with Extended Pedestrian A (EPA) channel model \cite{3GPP_epa}. The chosen pulse shape for the GFDM prototype filter is the raised cosine (RC) filter with a roll-off factor ($a$) of $0.5$. The active subcarrier indices are selected using the lookup table in Table \ref{tab:Look_up1}. Fine detector model parameters and summary are given in Table \ref{tab:model_params} and \ref{tab:model_summary}, respectively. During training stage, signal-to-noise ratio (SNR) is set to $15$dB, Adam optimizer \cite{Adam}, which is SGD-based, is used, and the learning rate is set to $8\times10^{-4}$. The DeepConvIM model is trained in a short time, $60$ epochs is enough to get significant results. A GFDM-IM training data set, including $16\times10^4$ symbols, and a GFDM-IM testing data set, including  $3\times10^4$ symbols are generated for each SNR value regarding $K=32$. For $K = 8$, training data and testing data include $32\times10^4$ and $6\times10^4$ GFDM-IM symbols, recpectively.

DeepConvIM model is constructed using Keras\cite{keras} (backend Tensorflow \cite{tensorflow}) and trained on Google Colab, providing tensor processing units (TPUs) in the cloud environment.

\begin{figure}[!h]
	\centering
	\vspace*{-0.4cm}
	\begin{center}{\includegraphics[scale = 0.4]{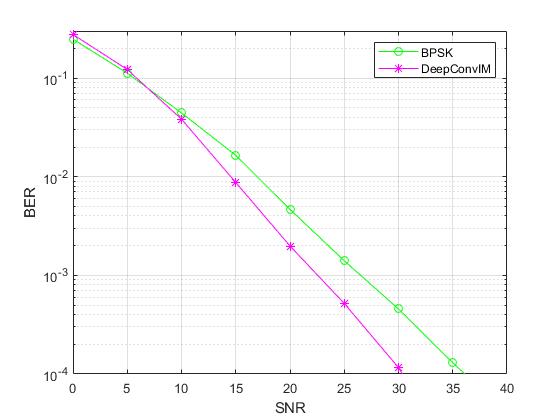}}
		\caption{BER performance of ZF and DeepConvIM with ZF coarse detector for BPSK transmission, ($K$ = 32, $M$ = 1)}\vspace*{-0.2cm}.
		\label{fig:bpsk}
	\end{center}
	\vspace*{-0.4cm}
\end{figure}

Fig. \ref{fig:bpsk} compares the BER performance of the ZF and the DeepConvIM with ZF coarse detector for binary phase shift keying (BPSK) transmission when $K = 32$ and $M = 1$. From Fig. \ref{fig:bpsk}, it is observed that the DeepConvIM provides approximately 6 dB better BER performance than ZF at a BER value of $10^{-4}$ .

\begin{figure}[!h]
	\centering
	\begin{center}{\includegraphics[scale = 0.4]{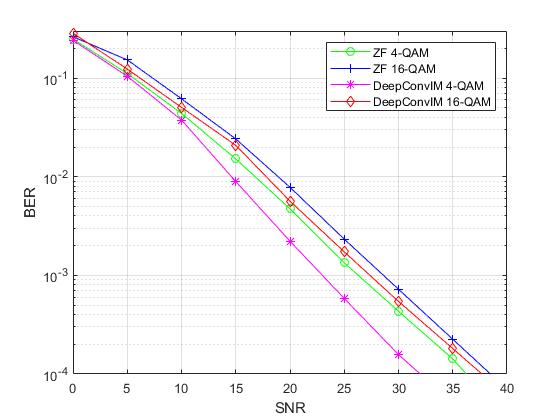}}
		\caption{BER performance of ZF and DeepConvIM with ZF coarse detector for 4-QAM and 16-QAM transmissions ($K$ = 32, $M$ = 1).}\vspace*{-0.2cm}
		\label{fig:qam}
	\end{center}
	\vspace*{-0.5cm}
\end{figure}

Fig. \ref{fig:qam} compares the BER performance of the ZF  and the DeepConvIM with ZF coarse detector  for 4-QAM and 16-QAM transmissions when $K = 32$ and $M = 1$. From Fig. \ref{fig:qam}, it is observed that the DeepConvIM provides approximately $4.5$ and $1$ dB better BER performance than ZF for 4-QAM and 16-QAM transmissions, respectively, at a BER value of $10^{-4}$.
	
\begin{figure}[!h]
	\centering
	\vspace*{-0.4cm}
	\begin{center}{\includegraphics[scale = 0.4]{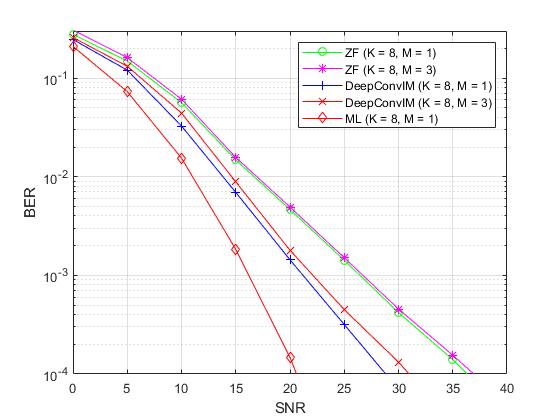}}
		\vspace*{-0.20cm}\caption{BER performance of ZF and DeepConvIM with ZF coarse detector for BPSK transmission ($K$ = 8, $M$ = (1 , 3)).}\vspace*{-0.20cm}
		\label{fig:bpsk_8}
	\end{center}
	\vspace*{-0.5cm}
\end{figure}

Fig. \ref{fig:bpsk_8} compares the BER performance of the ZF, ML and the DeepConvIM with ZF coarse detector for BPSK transmission, when $K = 8 $, $M = (1,3)$. From Fig. \ref{fig:bpsk_8}, at a BER value of $10^{-4}$, while DeepConvIM provides $6$ dB BER improvement with respect to ZF detector for $M = 3$, the BER improvement of  DeepConvIM with ZF coarse detector with respect to ZF detector is increased to $8$ dB when $M = 1$. On the other hand, ML detector has $8$ dB BER improvement with respect to DeepConvIM when $M = 1$.

From Fig. \ref{fig:bpsk} and \ref{fig:qam}, it is observed that as the modulation order increases, the learning capacity of the model decreases. From Fig. \ref{fig:qam}, it is observed that when the number of subsymbols increases, performance of the model decreases. On the other hand, from Fig. \ref{fig:bpsk} and \ref{fig:bpsk_8}, it is observed that when the number of subcarriers decreases, performance of the model increases. The number of CMs needed by the detectors in Fig. \ref{fig:bpsk}, \ref{fig:qam} and \ref{fig:bpsk_8} are given in Table \ref{tab:compucomp_all}. As mentioned earlier, DeepConvIM can be evaluated as a intermediate solution regarding computational complexity.

\section{Conclusion}
In this paper, a new GFDM-IM receiver scheme, which is constructed by the combination of a ZF detector and a deep convolutional neural network, has been proposed. BER performance of the proposed scheme has been compared to ZF and ML detectors by computer simulations under Rayleigh multipath fading channels. The proposed scheme has very simple and flexible neural network structure, which can be adapted to yield a trade-off between complexity and BER performance. It has been demonstrated that the proposed scheme provides significant BER improvement compared to ZF detector with a reasonable complexity increase. We conclude that deep convolutional learning-aided GFDM-IM scheme can be considered a promising PHY layer technique for beyond 5G wireless networks. As a future work, we will study application of deep learning to MIMO-GFDM systems.




%



\bibliographystyle{IEEEtran}
\bibliography{IEEEabrv,dl_aided_gfdm}

\end{document}